\documentclass[twocolumn,showpacs,preprintnumbers,showkeys,superscriptaddress]{revtex4}
\usepackage{graphicx}
\usepackage{dcolumn}
\usepackage{bm}

\def\eqref#1{Eq.~(\ref{eq:#1})}

\begin{document}

\title {$J$-pairing interaction, number of states,  and nine-$j$ sum rules of four identical particles}
\author{Y. M. Zhao}     \email{ymzhao@sjtu.edu.cn}
\affiliation{Department of Physics,  Shanghai Jiao Tong
University, Shanghai 200240, China}  \affiliation{Cyclotron
Center, Institute of Physical Chemical Research (RIKEN), Hirosawa
2-1, Wako-shi,  Saitama 351-0198,  Japan}\affiliation{Center of
Theoretical Nuclear Physics, National Laboratory of Heavy Ion
Accelerator, Lanzhou 730000, China}
\author{A. Arima} \affiliation{Science Museum,
Japan Science Foundation, 2-1 Kitanomaru-koen, Chiyodaku, Tokyo
102-0091, Japan}

\date{\today}

\begin{abstract}
In this paper we study $J$-pairing Hamiltonian and find that the
sum of  eigenvalues of spin $I$ states equals sum of norm matrix
elements within the pair basis for four identical particles such
as four fermions in a single-$j$ shell or four bosons with spin
$l$. We relate number of states to sum rules of nine-$j$
coefficients. We obtained sum rules for nine-$j$ coefficients
$\langle (jj)J,(jj)K:I| (jj)J, (jj)K:I\rangle$ and $\langle
(ll)J,(ll)K:I| (ll)J, (ll)K:I\rangle$ summing  over (1) even $J$
and even  $K$,  (2)  even $J$ and odd $K$, (3) odd $J$ and odd
$K$,  and (4) both even and odd values for $J$ and $K$, where $j$
is a half integer and  $l$ is an integer.
\end{abstract}


\maketitle

\section{Introduction}

The $J$-pairing Hamiltonian for a single-$j$ shell is an important
topic to study nuclear structure theory and also to study general
many-body systems. For the case of $J=0$, i.e., the monopole
pairing interaction, the famous seniority scheme
\cite{Racha,Talmi-book} provides exact solutions; for $J=J_{\rm
max}$, the  ``cluster" picture of Ref. \cite{Zhao-prc70} presents
an asymptotic classification of states. For other $J$ cases, it
was found that pairs with spin $J$ are reasonable building blocks
for low-lying states but little is known about exact eigenvalues
\cite{Zhao-prc68}.  In this paper  we shall go one step forward
along this line by proving that for four identical particles the
sum of eigenvalues for the $J$- pairing interaction is connected
with a sum of nine-$j$ coefficients.

The enumeration of  spin $I$ states (the number of spin $I$ states
is denoted by $D_I$ in this paper) for fermions in a single-$j$
shell or bosons with spin $l$ (We use a convention that $j$ is a
half integer and $l$ is an integer) is also a very common practice
in nuclear structure theory. $D_I$ is usually obtained by
subtracting the number of states with total angular momentum
projection $M=I+1$ from that with $M=I$ \cite{Lawson}. Because
numbers of states with different $M$'s seem irregular, $D_I$
values are usually tabulated in text-books, for sake of
convenience. Other methods include Racha's method \cite{Racha} in
terms of seniority scheme, generating function mehtod proposed and
studied by Katriel {\it et al.} \cite{Katriel}, Sunko and
collaborators \cite{Sunko}. All  these works are interesting and
important. However, the results are not algebraic. It is therefore
desirable to obtain analytical formulas of $D_I$. For $n=1$ and 2,
$D_I$ is known and is understood very well; but the situation
becomes complicated when $n\ge 3$, except for a few cases with
$I\sim I_{\rm max}$.

Historically, the first interesting formula of $D_I$ was given for
the case with $I=0$ and $n=4$  by Ginocchio and Haxton in Ref.
\cite{Ginocchio}. Their result  was revisited by Zamick and
Escuderos in Ref. \cite{Zamick}. In Ref. \cite{Zhao},  authors of
the present paper empirically constructed $D_I$ for $n=3$ and 4,
and some $D_I$'s for $n$=5. Recently, Talmi suggested a recursion
formula of $D_I$ in Ref. \cite{Talmi_PRC}, and used this formula
to prove $D_I$ formulas obtained empirically for $n=3$ in Ref.
\cite{Zhao}. Talmi's recursion formula is also readily applied to
prove the empirical formulas of Ref. \cite{Zhao} for $n=4$.  In
Ref. \cite{Zhao1}, we showed  that $D_I$ of $n$ particle systems
can be enumerated by using the reduction from SU($n+1$) to SO(3),
and as an example, $D_I$ for $n=4$ was obtained analytically.

The results of $D_I$ for three identical particles were applied to
obtain a number of sum rules for six-$j$ symbols in Appendix of
Ref. \cite{Zhao-prc70}. One can ask whether the results for $n=4$
can be used similarly to obtain sum rules for nine-$j$ symbols. If
the answer is {\it yes}, the application would be very
interesting, because sum rules of angular momentum couplings is
widely applied in many branches of physics, in particular in
nuclear structure theory (angular momentum coupling-recoupling
coefficients and sum rules were  compiled in Ref. \cite{Russians}
in 1988). This paper addresses the following question: can we
obtain sum rules for nine-$j$ symbols  based on studies of
$J$-pairing Hamiltonian and number of spin $I$ states for four
identical particles ? Furthermore, how far can one go along this
line?

This paper is organized as follows.  In Sec. II, we study
$J$-pairing Hamiltonian in order to obtain summation of all
non-zero eigenvalues in the presence of only one $J$-pairing
force. In Sec. III, we present sum rules of nine-$j$ symbols found
by using these summations  and  number of states for $n=4$
obtained in earlier works. In this paper we show that the  $D_I$
formulas provide us with a bridge between the $J$-pairing
interaction and sum rules of nine-$j$ symbols for identical
particles. The summary and discussion are given in Sec. IV.
Appendix A present formulas of nine-$j$ symbols in some special
cases. Appendix B discusses number of matrices involved in our sum
rule calculations.

\section{$J$-Paring interaction}

In this section we discuss the $J$-pairing interaction only for
identical fermions in a single-$j$ shell. Similar results are
readily obtained for  bosons with spin $l$.  Our $J$-pairing
Hamiltonian $H_J$  is defined as follows.
\begin{eqnarray}
&&  H_J = G_J \sum_{M = -J}^J A_M^{(J) \dagger} \ A_{M}^{(J) },  ~
A_M^{(J) \dagger} = \frac{1}{\sqrt{2}} \left[ a_{j}^{\dagger}
\times a_{j}^{\dagger}
     \right]^{(J)},
\nonumber \\
&&
     {A}_M^{(J)} = - (-1)^M\frac{1}{\sqrt{2}} \left[ \tilde{a}_{j} \times
     \tilde{a}_{j} \right]_{-M}^{(J)}, ~ ~ ~
     \nonumber \\
&&
     \tilde{A}^{(J)} = - \frac{1}{\sqrt{2}} \left[ \tilde{a}_{j} \times
     \tilde{a}_{j} \right]^{(J)},   \label{pair}
\label{H}
\end{eqnarray}
where $[~]_M^{(J)}$ means coupled to angular momentum $J$ and
projection $M$. We take $G_J=1$ in this paper.

For $n=3$, it was shown in Ref. \cite{Zhao-prc70} that there is
only one non-zero eigenvalue for $H_J$ when $I \ge j-1$, and all
eigenvalues are zero when $I < j-1$. For $n=4$  the situation is
more complicated, because there can be many non-zero eigenvalues
of spin $I$ states for $H_J$, and most of these eigenvalues are
not known except $I=0$ and $I\simeq I_{\rm max}$. However, their
summation is the trace of $H_J$ matrix with total spin $I$, and is
a constant with respect to any linear transformation. This trace
can be obtained  by summing the diagonal matrix elements
\begin{eqnarray}
\langle 0 |\left[ A^{(J)} \times A^{(K)} \right]^{(I)}_M  \left[
A^{(J)\dag} \times
A^{(K)\dag} \right]^{(I)}_M |0 \rangle   \nonumber   \\
=1 + (-)^I \delta_{JK}
                    - 4 (2J+1) (2K+1)
     \left\{%
     \begin{array}{ccc}
     j    & j  & J \\
     j    & j  & K \\
     J    & K  & I   \end{array}
     \right\} ~
\label{e5.12}
\end{eqnarray}
over $K$. Here $J$ and $K$ take only even values. This fact can be
proved by using two-body coefficients of fractional parentages
which are defined by
\begin{eqnarray}
&& \langle  \left. j^4 \alpha IM|  \right\}  j^2(J), j^2 (K) I
\rangle
 \nonumber \\
&=& \frac{1}{\sqrt{6}} (-)^{I} \langle  j^4 \alpha IM|\left[
A^{(J)\dag} \times A^{(K)\dag}  \right]^{(I)}_M |0 \rangle  ~.
\nonumber
\end{eqnarray}
The trace can be calculated as follows,
\begin{eqnarray}
 && \sum_{\alpha} \langle j^4  \alpha I | H_J | j^4  \alpha I
\rangle \nonumber \\
& = & \sum_K \sum_{\alpha} 6  \langle \left. j^2(J), j^2 (K) I |
\right\} j^4 \alpha I \rangle^2    \nonumber \\
& = & \sum_K \sum_{\alpha} \langle 0| \left[ A^{(J)}  \times
A^{(K)}
\right]^{(I)}_M |  I \alpha j^4   \rangle \nonumber \\
&& ~~~  \times \langle I \alpha j^4|\left[ A^{(J)\dag} \times
A^{(K)\dag}
\right]^{(I)}_M |0 \rangle \nonumber \\
 & = &  \sum_K \langle 0 |\left[ A^{(J)} \times A^{(K)} \right]^{(I)}_M  \left[
A^{(J)\dag} \times A^{(K)\dag} \right]^{(I)}_M |0 \rangle ~,
\nonumber
\end{eqnarray}
where $G_J=1$ is used. This is just the summation of Eq.
(\ref{e5.12}) over even $K$. One can also regard Eq. (\ref{e5.12})
as a simple generalization of the result in Ref.
\cite{Zhao-prc70}, where it was shown that the non-zero eigenvalue
of $H_J$ for spin $I$ states of three particles is given by the
norm $\langle j (j^2) J : I | j (j^2) J : I \rangle$. Note that
similar results are applicable to bosons with spin $l$.

Let us look at nine-$j$ symbols of identical particles under
certain conditions.  Based on Eq. (\ref{e5.12}) we easily find the
following well-known fact
\begin{eqnarray}
     \left\{%
     \begin{array}{ccc}
     j    & j  & J \\
     j    & j  & J \\
     K    & K'  & I   \end{array}
     \right\}  = 0~ \label{formula1}
\end{eqnarray}
for odd $I$ when $K \neq J$ or $K' \neq J$, based on the fact that
two identical pairs produce only even values of $I$ and thus the
norm in Eq. (\ref{e5.12}) equals zero. Here $K$ and $K'$ take even
values, or odd values simultaneously. This can be also seen from
the permutation symmetry of the nine-$j$ symbol, which requires
the left hand side of Eq. (\ref{formula1}) equals zero unless
$K+K'+I$ is even. We note without details that this formula is
also applicable to four bosons with spin $l$, i.e., one can
replace $j$ by $l$ in formula (\ref{formula1}). This is a
generalization of the result in Ref. \cite{Zamickx}, where it was
found
\begin{eqnarray}
     \left\{%
     \begin{array}{ccc}
     j    & j  & 2j-1 \\
     j    & j  & 2j-1 \\
     2j-1    & 2j-3  & 4j-4   \end{array}
     \right\}  = 0~.  \nonumber
\end{eqnarray}

The norm of Eq. (\ref{e5.12}) equals zero when $I=4j-7, 4j-5,
4j-4$, because there are no such states. We  find thus
\begin{eqnarray}
     \left\{%
     \begin{array}{ccc}
     j    & j  & J \\
     j    & j  & K \\
     J    & K  & I   \end{array}
     \right\}  = \frac{1}{4(2J+1)(2K+1)}~
\end{eqnarray}
for $I=4j-7, 4j-5, 4j-4$ and $J\neq K$ ($J,K$ are even). This is
also a generalization of a formula in Ref. \cite{Zamickx}:
\begin{eqnarray}
     \left\{%
     \begin{array}{ccc}
     j    & j  & 2j-3 \\
     j    & j  & 2j-1 \\
     2j-3    & 2j-1  & I   \end{array}
     \right\}  = \frac{1}{4(4j-5)(4j-1)}~ \nonumber
\end{eqnarray}
for $I=4j-7, 4j-5, 4j-4$.  Similarly, we have
\begin{eqnarray}
     \left\{%
     \begin{array}{ccc}
     j    & j  & 2j-1 \\
     j    & j  & 2j-1 \\
     2j-1    & 2j-1  & I   \end{array}
     \right\}  = \frac{1}{2(4j-1)^2}~
\end{eqnarray}
for $I=4j-2, 4j-4$. This formula was  also obtained in Ref.
\cite{Zamickx} and holds for both integer and half-integer value
of $j$. In Appendix A, we present some explicit formulas of
nine-$j$ symbols with $J=K=2j$ or $J=K=2j-1$. For completeness, we
also refer Refs. \cite{Zhao-prc68,Russians,Zamickx,Zamickx1},
concerning formulas of six-$j$ and nine-$j$ symbols for identical
particles.

Now we enumerate the number of matrices of Eq. (\ref{e5.12}) with
different $J$. This is related to the number of non-zero two-body
coefficients of fractional parentage which was  obtained for
specific examples in studying regularities of energy centroids in
the presence of random interactions \cite{E-average}. Without
going to details we present the results of the number of matrices
involved in Eq.(\ref{e5.12}) with different $J$ as below.

For $I \ge 2j$, the number of matrices with $K=J$ is given by
\begin{equation}
\left[ (4j+2- I)/4\right] \label{e5.13}
\end{equation}
and the number of matrices with $K \neq J$ is
\begin{equation}
\left[ (4j-I)/2 \right] \left( \left[ (4j-I)/2 \right] +1
\right)/2 - \left[ (4j+2- I)/4\right] ~.
 \label{e5.14}
\end{equation}
The $\left[ ~~\right]$ in this paper means to take the largest
integer not exceeding the value inside.

For $I \le 2j-1$, the number of matrices with $J=K$ is always 1
for even $I$, the case with $J \neq K$ is more complicated and the
number of such matrices is given in the Appendix B. It is noted
that $J$ and $K$ take only even values in this Section.

\section{Sum rules for nine-j symbols}

The procedure to obtain sum rules of nine-$j$ symbols in this
paper is straightforward. In Sec. II we obtain summation of
eigenvalues for $H=H_J$. From the sum rule of two-body
coefficients of fractional parentage, one obtains
$\frac{n(n-1)}{2}$ multiplied by the number of spin $I$ states,
$D_I$, if one sums Eq. (\ref{e5.12}) over even $J$ and even $K$,
namely:
\begin{eqnarray}
& & \sum_J \sum_{\alpha} \langle j^4  \alpha I | H_J | j^4  \alpha
I \rangle \nonumber \\
&=& \sum_{{\rm even ~} J ~ {\rm even ~} K }  \langle 0 |\left[
A^{(J)} \times A^{(K)} \right]^{(I)}_M  \left[ A^{(J)\dag} \times
A^{(K)\dag}
\right]^{(I)}_M |0 \rangle  \nonumber \\
&  =&  6 D_I ~. \label{sum-1}
\end{eqnarray}
where  $D_I$ formulas were given in Refs. \cite{Zhao,Zhao1}. New
sum rules of nine-$j$ symbols now can be obtained by using $D_I$
formulas, Eq. (\ref{e5.12}) and Eq. (\ref{sum-1}).

For realistic systems both $J$ and $K$ are even, as in
Eq.(\ref{e5.12}) of Sec. II and Eq. (\ref{sum-1}). In this paper
we also discuss sum rules of nine-$j$ symbols under other
conditions for  $J$ and $K$, such as odd $J$ and odd $K$, etc. We
denote
\begin{eqnarray}
&& S_I (j^4, {\rm condition ~ X~ on }~ J {\rm ~and ~}K)
\nonumber \\
&& =  \sum_{{\rm X  }  } 4 (2J+1) (2K+1)
     \left\{
     \begin{array}{ccc}
     j    & j  & J \\
     j    & j  & K \\
     J    & K  & I   \end{array}
     \right\}   \label{condition}
\end{eqnarray}
for sake of simplicity.  The condition $X$ of the sum rules for
$J$ and $K$ will be one of the following: (1) even $J$ and even
$K$ (realistic); (2) even $J$ and odd $K$ or odd $J$ and even $K$;
(3) odd $J$ and odd $K$; and (4) both even and odd values for $J$
and $K$. Conditions (2-4) are not  physical for identical
particles in quantum mechanics. We similarly define $S_I (l^4,
{\rm condition ~ X~ on }~ J {\rm ~and ~}K)$ for $l$.

First we present our results of $S_I (j^4, $ requirement $X$ on
$J$  and $K$), which provides us with rich sum rules of nine-$j$
symbols.

For $I \ge 2j$, one obtains
\begin{eqnarray}
&& S_I(j^4, {\rm even ~} J ~ {\rm even ~} K)
    \nonumber \\
 &&    =  \frac{1}{2}
     \left[ \frac{4j   - I}{2} \right] \times \left[ \frac{4j   - I +2}{2} \right]
   \nonumber \\
 && (-)^I \left[ \frac{4j +2 - I}{4} \right] - 6D_I    ~,
 \label{sumx0}
\end{eqnarray}
based on Eqs. (\ref{e5.12}), (\ref{e5.13}),(\ref{e5.14}), and
(\ref{sum-1}).  Let us introduce $I_0$ by the relation   $I =
I_{\rm max} - 2 I_0$ for even $I$ and $I = I_{\rm max} - 2I_0 -3$
for odd $I$, where $I_{\rm max} = 4j-6$.  Using the $I_0$  we can
rewrite $D_I= D_{I_{\rm max} - 2 I_0}$ for even $I$ and $D_I  =
D_{I_{\rm max} - 2 I_0-3}$ for odd $I$.  According to Ref.
\cite{Zhao1},
\begin{eqnarray}
&& D_I = 3 \left[ \frac{I_0}{6} \right] (\left[ \frac{I_0}{6}
\right]+1)
- \left[ \frac{I_0}{6} \right] \nonumber \\
& + & (\left[ \frac{I_0}{6} \right] + 1) \left( (I_0 {\rm ~ mod ~}
6) +1 \right) + \delta_{(I_0 {\rm ~ mod ~} 6), 0} -1 ~. \nonumber
\end{eqnarray}
 One thus has
\begin{eqnarray}
&& S_I(j^4, {\rm even ~} J ~ {\rm even ~} K ) \nonumber \\
&&=  (-)^I \left[ \frac{4j +2 - I}{4} \right]
 + \frac{1}{2}
     \left[ \frac{4j   - I}{2} \right] \times \left[ \frac{4j   - I +2}{2} \right]
 \nonumber \\
&& - 18 \left[ \frac{I_0}{6} \right] \left( \left[ \frac{I_0}{6}
\right]+1 \right)
+ 6 \left( \left[ \frac{I_0}{6} \right] +1 \right) \nonumber \\
&  - & 6(\left[ \frac{I_0}{6} \right] + 1) \left( (I_0 {\rm ~ mod
~}
6) +1 \right) - 6\delta_{(I_0 {\rm ~ mod ~} 6), 0}  ~. \nonumber \\
\label{sumx1}
\end{eqnarray}
The behavior of the right hand side is not easy to see due to
those $(I_0 {\rm~ mod ~ }6)$, $\delta$ and $\left[ \frac{I_0}{6}
\right]$, and so on.  The situation becomes much more transparent
when one writes $S_I(j^4$, even $J$ even $K$) values explicitly.
For $I=$even, we find the following formulas:
\begin{eqnarray}
&& S_I(j^4, {\rm even ~} J ~ {\rm even ~} K)   = \left\{
\begin{array}{ll}
2 & {\rm for ~} I = I_{\rm max} , \\
6 & {\rm for ~} I = I_{\rm max} -2 , \\
6 & {\rm for ~}  I =I_{\rm max}  -4 ,\\
6 & {\rm for ~}  I =I_{\rm max} -6 , \\
8 & {\rm for ~}  I =I_{\rm max} -8 , \\
10 &  {\rm for ~} I=I_{\rm max} -10 , \\
\vdots &  ~~~~  ~~~~~~~~~~~ \vdots
\end{array} \right. ~~~;  \nonumber \\
 \label{sumx2}
\end{eqnarray}
for $I=$odd we can use Eq. (\ref{sumx2}) to obtain the sum rules:
$S_I(j^4,$ even $J$ even $K)= S_{I+3} (j^4$, even $J$ even $K$).
We find that $S_I(j^4, $ even $J$ even $K$) has a modular
behavior:
\begin{eqnarray}
&& S_I(j^4, {\rm even ~} J ~ {\rm even ~} K) \nonumber \\
& =& S_{((I_{\rm max}-I) {\rm ~ mod~} 12)}(j^4, {\rm even ~} J ~
{\rm even ~} K) + 6 \left[\frac{I_{\rm max}-I}{12} \right] ~.
\nonumber \\
\end{eqnarray}
For $I = I_{\rm max} -1$, one obtains $S_I(j^4, {\rm even ~} JK) =
4$ based on the right hand side of Eqs. (\ref{e5.12}) and
(\ref{sum-1}).

 For $I \le 2j-1$, Eq. (\ref{sum-1}) is less transparent to be
simplified \footnote{One could write a lengthy sum rule in this
case based on $D_I$ of Ref. \cite{Zhao1} and a rather complicated
formula for number of $J=K$ and $J\neq K$ matrices of Eq.
(\ref{e5.12})}, due to the complexity for $D_I$ formula (See Eq.
(3) of Ref. \cite{Zhao}) and number of $J=K$ and $J\neq K$
matrices of Eq. (\ref{e5.12}). However,  by using Eq.(\ref{sum-1})
of this paper, Eq. (3) of Ref. \cite{Zhao}, and results in
Appendix B, one is able to obtain explicitly the sum rules for $I
\le 2j-1$:
\begin{eqnarray}
&& S_I(j^4, {\rm even ~} J ~ {\rm even ~} K) \nonumber \\
&=&  \left\{
\begin{array}{ll}
 2  m -2 & {\rm for ~} I = 0  ,\\
 0  & {\rm for ~} I = 1 \\
 4 - 2m    &  {\rm for ~} I =2 , \\
 2  m  & {\rm for ~} I = 3 , \\
 2 &  {\rm for ~} I =4  \\
 4 - 2 m & {\rm for ~} I = 5 , \\
 2 + 2 m & {\rm for ~}  I =6 , \\
 4 & {\rm for ~} I = 7  ,\\
 6- 2 m & {\rm for ~}  I =8 , \\
 2 + 2 m & {\rm for ~} I = 9  ,\\
 6 &    {\rm for ~} I =10 , \\
8 - 2 m &    {\rm for ~} I =11 , \\
~~~~~~  \vdots &    ~~~~~~~\vdots
\end{array} \right. \nonumber \\  \label{Sum-x2}
\end{eqnarray}
has a modular behavior: \begin{eqnarray}
&& S_I(j^4, {\rm even ~} J ~ {\rm even ~} K) \nonumber \\
 &=& S_{(I {\rm ~ mod} ~ 12)} (j^4, {\rm even ~} J ~ {\rm even ~} K)
+ 6 \left[\frac{I}{12} \right] ~. \label{sum-eex}
\end{eqnarray}
In Eq.(\ref{Sum-x2}), $m=(j+3/2) ~{\rm mod} ~3$.

 In Eqs.(\ref{sum-1}) and (\ref{sumx0}-\ref{sum-eex}) $J$
and $K$ take only even values. It is interesting to discuss
whether there are simple sum rules in which $J$ and $K$ can be
both even and odd. For this case $0\le I \le I_{\rm max} = 4j$.
Starting from Eq. (9.29) of Ref. \cite{Talmi-book} for
$J_1=J_2=J_3=J_4$, $J_{12}=J_{13}=J$, $J_{34}= J_{24}=K$, $J=I$,
we multiply $4(2J+1)(2K+1)$ and sum over all $JK$ (i.e., $J$ and
$K$ take both even and odd non-negative integers). Using Eq.
(9.28) of Ref. \cite{Talmi-book}, we find
\begin{eqnarray}
&&  S_I(j^4, {\rm both ~ even ~ and ~ odd ~ values ~ for ~} J {\rm
~and~} K) \nonumber \\
&& = \sum_{J,K=0; \Delta(JKI)}^{2j} (-)^{J+1} \nonumber \\
&&  = \left\{
\begin{array}{ll}
 4  \left[ (1+I)/2  \right]& {\rm for ~} I \le 2j+1  ,\\
4 +4 \left[ (4j - I)/2  \right]    &  {\rm for ~} I \ge 2j  ~,
\end{array} \right.  \label{all-j}
\end{eqnarray}
where $\Delta(JKI)$ means that $J,K$ and $I$ satisfy the triangle
relation of angular momentum coupling.

If both $J$ and $K$ are odd values, $0 \le I \le I_{\rm max} =
4j$. In this case we consider fictitious (not realistic for
identical particles) ``bosons" with a half integer spin $j$.
According to Ref. \cite{Zhao1}, the number of states $D_I$ for
four bosons with spin $j$ equals that for four fermions in a
single-$l$ shell with $2l=2j+3$. As $D_I$ for four fermions in a
single-$l$ shell was given in Ref. \cite{Zhao-prc68}, we can
derive $S_I(j^4$, odd $J$ odd $K$) by using Eqs. (\ref{e5.12}) and
(\ref{sum-1}), together with Eqs. (3-5) in Ref. \cite{Zhao-prc68}.
Similar to Eqs. (\ref{sumx0}),(\ref{sumx1}) and (\ref{Sum-x2}), we
obtain that when $I \le 2j$,
\begin{eqnarray}
&&  S_I(j^4, {\rm odd ~} J ~ {\rm odd~}K) \nonumber \\
&& = \left\{
\begin{array}{ll}
 2 -   2 m  & {\rm for ~} I = 0  ,\\
  0 & {\rm for ~} I = 1  ,\\
 2 m    &  {\rm for ~} I =2 , \\
 4-2 m    & {\rm for ~} I = 3 , \\
 2 &  {\rm for ~} I =4  \\
 2 m &  {\rm for ~} I = 5 , \\
6  - 2 m   &  {\rm for ~} I =6 , \\
 4 & {\rm for ~} I = 7  ,\\
2+  2 m  & {\rm for ~}  I =8 , \\
6-2 m & {\rm for ~} I = 9  ,\\
6 &   {\rm for ~} I =10 , \\
4+ 2 m &  {\rm for ~}  I =11 , \\
 ~~~~ \vdots &   ~~~~~~~ \vdots
\end{array} \right. \nonumber \\ \label{SSS}
\end{eqnarray}
has a modular behavior:
\begin{eqnarray}
&& S_I (j^4, {\rm odd ~} J ~ {\rm odd~}K) \nonumber \\
&=& S_{\left( I {\rm ~ mod ~} 12 \right)}(j^4, {\rm odd ~} J ~
{\rm odd~}K) +6 \left[\frac{I}{12} \right] ~.
\end{eqnarray}
where $m =  (j-3/2) {\rm ~ mod~} 3 $ in Eq. (\ref{SSS}); and when
$I \ge 2j$
\begin{eqnarray}
&&  S_I(j^4, {\rm odd ~} J ~ {\rm odd~}K) \nonumber \\
=&&  \left\{
\begin{array}{ll}
4  & {\rm for ~} I = 4j   ,\\
2 & {\rm for ~} I = 4j -2  ,\\
4    &  {\rm for ~}I = 4j -4 , \\
6    & {\rm for ~} I=4j - 6 , \\
6 &  {\rm for ~} I=4j -8  \\
6 &  {\rm for ~} I=4j - 10 , \\
\vdots &    ~~~~~~~~ \vdots
\end{array} \right.
\end{eqnarray}
has a modular behavior:
\begin{eqnarray}
&& S_I(j^4, {\rm odd ~} J ~ {\rm odd~}K) \nonumber \\
&& = S_{( {(4j-I)} {\rm ~ mod ~} 12 )} (j^4, {\rm odd ~} J ~ {\rm
odd~}K) + 6 \left[\frac{4j-I}{12} \right] \nonumber \\
\end{eqnarray}
for even $I$, and $S_I(j^4$, odd $J$ odd $K$) = $S_{I+3}(j^4$, odd
$J$ odd $K$) for odd $I$. For $I=4j -1$ (odd $I$), $S_I(j^4$, odd
$J$ odd $K$) = 0.

For even $J$ and odd $K$ or for odd $J$ and even $K$, $0 \le I \le
I_{\rm max} = 4j-1$. For this case
\begin{eqnarray}
&& S_I(j^4,  {\rm ~even ~} J ~ {\rm odd ~} K  ) \equiv S_I(j^4,
{\rm ~odd ~} J  ~ {\rm even ~} K  ) \nonumber \\
&&= \left( S_I(j^4,  {\rm ~both ~even ~ and ~odd ~ values ~ for~}
J {\rm ~and~} K ) \right. \nonumber \\
&&-\left. S_I(j^4, {\rm ~even ~} J {\rm ~even~} K )    - S_I(j^4,
{\rm ~ odd ~} J  ~ {\rm odd ~} K )\right)/2 ~.
 \nonumber
\end{eqnarray}
Using this relation and above results we find that when $I \le
2j$,
\begin{eqnarray}
&& S_I(j^4,  {\rm ~even ~} J ~ {\rm odd ~} K  ) \equiv S_I(j^4,
{\rm ~odd ~} J  ~ {\rm even ~} K  ) \nonumber \\
&&  = \left\{
\begin{array}{ll}
2\left[ \frac{I}{4} \right]  & {\rm for ~even ~} I    \\
2+2\left[ \frac{I}{4} \right]  & {\rm for ~odd ~} I
\end{array} \right. ~;
\end{eqnarray}
and when $I \ge 2j$,
\begin{eqnarray}
&& S_I(j^4,  {\rm ~even ~} J ~ {\rm odd ~} K  ) \equiv S_I(j^4,
{\rm ~odd ~} J  ~ {\rm even ~} K  ) \nonumber \\
&& = 2 + \left[ \frac{I_{\rm max} - I }{4} \right] ~.
\end{eqnarray}

\vspace{0.4in}

 Similarly, we obtain sum rules by replacing the half
integer $j$ to the integer $l$. First, let us study  the case for
even values of $J$ and $K$. We find that when $I\le 2l$,
\begin{eqnarray}
&& S_I(l^4, {\rm even ~ } J {\rm ~ even~} K) \nonumber \\
 &&=   \sum_{{\rm even ~ } J {\rm ~ even~} K} 4 (2J+1) (2K+1)
     \left\{
     \begin{array}{ccc}
     l    & l  & J \\
     l    & l  & K \\
     J    & K  & I   \end{array}
     \right\}    \nonumber \\
 && = \left\{
\begin{array}{ll}
4 - 2m & {\rm for ~} I = 0 , \\
0 & {\rm for ~} I = 1 , \\
2m & {\rm for ~} I = 2 , \\
2-2m   & {\rm for ~}  I = 3 , \\
4 & {\rm for ~}  I =4 ,\\
2m & {\rm for ~}  I = 5 , \\
6 -2m& {\rm for ~}  I =6 , \\
2 & {\rm for ~} I = 7 , \\
4+2m & {\rm for ~}  I =8 , \\
6-2m & {\rm for ~} I = 9 , \\
6 &  {\rm for ~} I=10 , \\
2+2m &  {\rm for ~} I= 11 , \\
\vdots &  ~~~~~~~~~~\vdots
\end{array} \right. \nonumber \\
\label{Sumx2}
\end{eqnarray}
has a modular behavior:
\begin{eqnarray}
&& S_I(l^4, {\rm even ~ } J {\rm ~ even~} K)  \nonumber \\
&& =S_{\left( I {\rm ~ mod ~} 12\right)}(l^4, {\rm even ~} J ~
{\rm even ~}K) + 6 \left[\frac{I_{\rm max}-I}{12} \right] ~,
\nonumber \\
\end{eqnarray}
where $m=l {\rm ~ mod ~} 3$ in Eq.(\ref{Sumx2}),  and when $I\ge
2l$,
\begin{eqnarray}
&&   \sum_{{\rm even ~ } JK} 4 (2J+1) (2K+1)
     \left\{
     \begin{array}{ccc}
     l    & l  & J \\
     l    & l  & K \\
     J    & K  & I   \end{array}
     \right\}    \nonumber \\
&&= \sum_{{\rm even ~} JK}   \left( 1 + (-)^I \delta_{JK} \right) \nonumber \\
&&  -18 \left[ \frac{I_0}{6} \right] (\left[ \frac{I_0}{6}
\right]+1)
+ 6 \left[ \frac{I_0}{6} \right] \nonumber \\
& - & 6(\left[ \frac{I_0}{6} \right] + 1) \left( (I_0 {\rm ~ mod
~}
6) +1 \right) - 6\delta_{(I_0 {\rm ~ mod ~} 6), 0} +6  \nonumber \\
 && = \left\{
\begin{array}{ll}
4 & {\rm for ~} I = I_{\rm max} , \\
0 & {\rm for ~} I = I_{\rm max} -1 , \\
2 & {\rm for ~} I = I_{\rm max} -2 , \\
4   & {\rm for ~}  I = I_{\rm max} -3 , \\
4 & {\rm for ~}  I =I_{\rm max}  -4 ,\\
2 & {\rm for ~}  I = I_{\rm max} -5 , \\
6 & {\rm for ~}  I =I_{\rm max} -6 , \\
4 & {\rm for ~} I = I_{\rm max} -7 , \\
6 & {\rm for ~}  I =I_{\rm max} -8 , \\
6 & {\rm for ~} I = I_{\rm max} -9 , \\
6 &  {\rm for ~} I=I_{\rm max} -10 , \\
6 &  {\rm for ~} I=I_{\rm max} -11 , \\
\vdots &  ~~~~~~~~~~\vdots
\end{array} \right.
\end{eqnarray}
has a modular behavior:
\begin{eqnarray}
&& S_I(l^4, {\rm even ~ } J {\rm ~ even~} K) \nonumber \\
&=& S_{\left( {(I_{\rm max}- I)} {\rm ~ mod ~} 12\right)}(l^4,
{\rm even ~} JK) +  6 \left[\frac{I_{\rm max}-I}{12} \right] ~,
\nonumber \\
\end{eqnarray}
where $I_{\rm max} = 4l$.

If $J$ and $K$ take both even and odd values, similar to the
process of obtaining Eq. (\ref{all-j}), we find for $I \le 2l$
\begin{eqnarray}
&&  S_I(l^4,  {\rm both ~ even ~ and ~ odd ~ values ~ for ~} J
{\rm
~and~} K) \nonumber \\
&&  = \left\{
\begin{array}{ll}
4+ 4  \left[ \frac{I}{2}  \right]& {\rm for ~even ~} I  ,\\
4 \left[ \frac{I}{2}  \right]    &  {\rm for ~odd ~} I
\end{array} \right. ~;
\end{eqnarray}
for $I \ge 2l$,
\begin{eqnarray}
&&  S_I(l^4,  {\rm both ~ even ~ and ~ odd ~ values ~ for ~} J
{\rm ~and~} K) \nonumber \\
&&  =  4 +4  \left[ \frac{4l - I}{2}  \right] ~.
\end{eqnarray}
We note this sum rule has the same form as Eq. (\ref{all-j}) for
$I\ge 2j$.

For odd $J$ and odd $K$ values, $0 \le I \le I_{\rm max} =4l-2$.
We find that when $I \le 2l$,
\begin{eqnarray}
&&  S_I(l^4, {\rm  odd~ } J {\rm ~odd ~} K)  \nonumber \\
&& = \left\{
\begin{array}{ll}
 2 m  & {\rm for ~} I = 0  ,\\
  0 & {\rm for ~} I = 1  ,\\
 4- 2 m   &  {\rm for ~} I =2 , \\
 -2  + 2m   & {\rm for ~} I = 3 , \\
 4 &  {\rm for ~} I =4  \\
 4- 2 m &  {\rm for ~} I = 5 , \\
2+ 2 m   &  {\rm for ~} I =6 , \\
 2 & {\rm for ~} I = 7  ,\\
 8- 2 m  & {\rm for ~}  I =8 , \\
 2 + 2m & {\rm for ~} I = 9  ,\\
6 &   {\rm for ~} I =10 , \\
 6- 2 m &  {\rm for ~}  I =11 , \\
 ~~~~~~\vdots &  ~~~~~~\vdots  \nonumber
\end{array} \right. \\ \label{SS1}
\end{eqnarray}
has a modular behavior:
\begin{eqnarray}
&& S_I (l^4, {\rm  odd~ } J {\rm ~odd ~} K) \nonumber \\
& = & S_{\left( I {\rm ~ mod ~} 12 \right)}(l^4, {\rm  odd~ } J
{\rm ~odd ~} K) + 6 \left[\frac{I}{12} \right]  ~,
\end{eqnarray}
where $m = l {\rm ~ mod ~} 3$ in Eq. (\ref{SS1}); and when $I \ge
2l$,
\begin{eqnarray}
&&  S_I(l^4, {\rm  odd~ } J {\rm ~odd ~} K)  \nonumber \\
&& = \left\{
\begin{array}{ll}
2   &  {\rm for ~} I_{\rm max}-I =0 , \\
 4 &  {\rm for ~} I_{\rm max}-I =2  \\
2 &  {\rm for ~} I_{\rm max}-I = 4 , \\
6   &  {\rm for ~} I_{\rm max}-I =6 , \\
6 & {\rm for ~} I_{\rm max}-I = 8 \\
6 & {\rm for ~} I_{\rm max}-I = 10 \\
\vdots &  ~~~~  ~~ ~~~\vdots
\end{array} \right.
\end{eqnarray}
has a modular behavior
\begin{eqnarray}
&& S_I (l^4,  {\rm  odd~ } J {\rm ~odd ~} K) \nonumber \\
& = &  S_{ ((I_{\rm max} - I) {\rm ~ mod ~} 12 )} (l^4, {\rm  odd~
} J {\rm ~odd ~} K) +  6 \left[\frac{I_{\rm max}-I}{12} \right] ~
; \nonumber \\
\end{eqnarray}
 For odd $I\ge 2l$, $S_I = S_{I+3}$ with $S_{I_{\rm max}-1} = 0$.

If we take odd $J$ and even $K$ values or we take even $J$ and odd
$K$ values, $0 \le I \le I_{\rm max} = 4l-1$. For this case
\begin{eqnarray}
&& S_I(l^4,  {\rm ~even ~} J ~ {\rm odd ~} K  ) \equiv S_I(l^4,
{\rm ~odd ~} J  ~ {\rm even ~} K  ) \nonumber \\
&&= \left( S_I(l^4,  {\rm ~both ~even ~ and ~odd ~ values ~ for~}
J {\rm ~and~} K ) \right. \nonumber \\
&&-\left. S_I(l^4, {\rm ~even ~} J {\rm ~even~} K )    - S_I(l^4,
{\rm ~ odd ~} J  ~ {\rm odd ~} K )\right)/2
~  \nonumber \\
 && = \left\{
\begin{array}{ll}
 2\left[  \frac{I+2}{4} \right] & {\rm for   ~} I \le 2l ~,  \\
 2+\left[  \frac{I_{\rm max}-I}{4} \right] & {\rm for ~} I \ge 2l
 ~. \end{array} \right.
\end{eqnarray}

\section{Summary and Discussion}

To summarize, in this paper we first show that  the sum of
eigenvalues of spin $I$ states for $J$-pairing interaction is
given by
\begin{eqnarray}
\sum_K \left(1 + (-)^I \delta_{JK} \right)
                    - 4 \sum_K (2J+1) (2K+1)
     \left\{%
     \begin{array}{ccc}
     j    & j  & J \\
     j    & j  & K \\
     J    & K  & I   \end{array}
     \right\} ~    \nonumber
\end{eqnarray}
for fermions, and
\begin{eqnarray}
\sum_K \left(1 + (-)^I \delta_{JK} \right)
                    - 4 \sum_K (2J+1) (2K+1)
     \left\{%
     \begin{array}{ccc}
     l    & l  & J \\
     l    & l  & K \\
     J    & K  & I   \end{array}
     \right\} ~    \nonumber
\end{eqnarray}
for bosons. Then we relate them with number of spin $I$ states to
obtain nine-$j$ sum rules. We study
\begin{eqnarray}
&&  4 (2J+1) (2K+1)
     \left\{
     \begin{array}{ccc}
     j    & j  & J \\
     j    & j  & K \\
     J    & K  & I   \end{array}
     \right\}    \nonumber
\end{eqnarray}
and
\begin{eqnarray}
 &&  4 (2J+1) (2K+1)
     \left\{
     \begin{array}{ccc}
     l    & l  & J \\
     l    & l  & K \\
     J    & K  & I   \end{array}
     \right\}    \nonumber
\end{eqnarray}
summing over $J$ and $K$ under following situations: (1) all $J$
and $K$ are even; (2) $J$ and $K$ can be both even and odd; (3)
all $J$ and $K$ are odd; (4) $J$ is even and $K$ is odd. We also
obtain formulas for special $J, K$ and $I$ values, based on the
physical meaning of the norm in Eq. (\ref{e5.12}).

Sum rules in Eqs. (A1-A2) of Ref. \cite{Zhao-prc70} can be
obtained as a special case of the results in this paper: $I=0$ for
Eqs. (\ref{Sum-x2}) and (\ref{Sumx2}). This work is therefore a
generalization of some of our earlier results. We use $J$ pairing
interaction as a tool to obtain the sum rules but these results
are independent of the interaction.

In Ref. \cite{Zhao1}, it was found that number of spin $I$ states
$D_I$  for four bosons with spin $l$ and that for four fermions in
a single-$j$ shell are the same when $2l = 2j-3$. This produce the
same value of the right hand side in Eq. (\ref{sum-1}) for
fermions and bosons. Unfortunately, number of $J$ and $K$ for
these two cases are different (number of $J$ for bosons is
$l+1$=$j-1/2$ while that for fermions is $j+1/2$), which present
different sum rules of the case with even values for both $J$ and
$K$.

One may ask how far one can go along this line, i.e., to construct
sum rules of angular momentum coupling by using formulas of $D_I$.
As $n$ increases, $D_I$ formulas and sum of eigenvalues of spin
$I$ states become more and more complicated. The situation is
already complicated for $n=4$. For $n=5$ there are $D_I$ formulas
for only $I\sim 0$ or $\sim I_{\rm max}$. Therefore, it is
difficult to obtain $D_I$ formulas and new sum rules of angular
momentum couplings in which more particles ($n\ge 5$) are
involved, except for a few cases with $I \sim I_{\rm max}$ where
the $D_I$ is given by a fixed number series \cite{Zhao,Talmi_PRC}.

Acknowledgement:  We would like to thank Prof. Igal Talmi for his
reading of our paper.  One of the authors (YMZ) would like to
thank the National Natural Science Foundation of China for
supporting this work under Grant Nos. 10545001 and 10575070.

%\newpage

\vspace{0.3in}

\begin{center}
{\bf Appendix ~A ~~~ Formulas of special nine-$j$ symbols}
\end{center}

In this Appendix we present formulas for  nine-$j$ symbol
\begin{equation}
    \left\{
     \begin{array}{ccc}
     j    & j  & J \\
     j    & j  & J \\
     J    & J  & I  \end{array}
     \right\} ~, \label{bbb1}
\end{equation}
where $J=2j$ or $2j-1$,  based on its expansion in terms of
six-$j$ symbols.  Value of $j$ in this Appendix can be either a
half integer or an integer. One sees that the nine-$j$ symbol of
Eq. (\ref{bbb1}) equals zero if $I$ is odd, because there appears
$(-)^{4j+4J+I}=(-)^I$ phase factor if one exchanges  the first and
the second row in Eq.(\ref{bbb1}). From this one obtains  that the
nine-$j$ symbol of Eq. (\ref{bbb1}) vanishes unless $I$ is even.
Below we discuss nine-$j$ symbols of Eq. (\ref{bbb1}), with $I$
being even and $J=2j$ or $2j-1$.

We define
\begin{equation}
  f'_m =   \left\{
     \begin{array}{ccc}
     j    & j  & 2j \\
     j    & j  & 2j \\
     2j    & 2j  & 4j-m  \end{array}
     \right\}  \nonumber
\end{equation}
and obtain  following formulas:
\begin{eqnarray}
&& f'_0 = \frac{1}{(4j+1)^2}  ~,\nonumber \\
&& f'_2 = \frac{-1}{2(4j+1)^2(4j-1)} ~,  \nonumber \\
&& f'_4 = \frac{3(2j-1)}{2(16j^2-1)^2(4j-3)} ~,  \nonumber \\
&& f'_6 = \frac{-3\times 5 (2j-2)}{4(4j-5)(4j-3)(4j-1)^2(4j+1)^2} ~, \nonumber \\
&& f'_8 = \frac{3\times 5 \times 7(2j-2)(2j-3)}{4(4j-7)(4j-5)(4j-3)^2(4j-1)^2(4j+1)^2} ~,\nonumber \\
&& f'_{10} = \frac{-3\times 5 \times 7\times 9}{8(4j-3)^2(4j-1)^2(4j+1)^2}  \nonumber \\
&&    ~~~~~~~~~\times \frac{(2j-4)(2j-3)}{(4j-9)(4j-7)(4j-5)}~,
\nonumber  \\
&& f'_{12} =  \frac{3\times 5 \times 7\times 9 \times 11
}{8(4j-5)^2(4j-3)^2(4j-1)^2(4j+1)^2)}  \nonumber \\
&&    ~~~~~~~~~\times
\frac{(2j-3)(2j-4)(2j-5)}{(4j-11)(4j-9)(4j-7)} ~. \nonumber
\end{eqnarray}

We define
\begin{eqnarray}
f_I =  \left\{
     \begin{array}{ccc}
     j    & j  & 2j \\
     j    & j  & 2j \\
     2j    & 2j  & I  \end{array}
     \right\}  \nonumber
\end{eqnarray}
and obtain  following formulas:
\begin{eqnarray}
&& f_0  = (-)^{2j} \frac{\left[ (2j-1)!
\right]^2}{(4j+1)^2(4j-1)!} \frac{1}{2} (2j)
 ~, \nonumber \\
&& f_2  = -(-)^{2j} \frac{\left[ (2j-1)!
\right]^2}{(4j+1)^2(4j-1)!} \frac{1}{2} \frac{(2j)(2j+1)}{(4j-1)}
 ~, \nonumber \\
&& f_4  =(-)^{2j} \frac{ \left[ (2j-1)! \right]}{(4j+1)^2(4j-1)!}
\frac{3}{4} \frac{(2j)(2j+1)(2j+2)}{(4j-3)(4j-1)}
 ~, \nonumber \\
&& f_6  =-(-)^{2j} \frac{ \left[ (2j-1)! \right]}{(4j+1)^2(4j-1)!}
\frac{5}{4}
\nonumber \\
&& ~~~ \times
\frac{(2j)(2j+1)(2j+2)(2j+3)}{(4j-5)(4j-3)(4j-1)} ~, \nonumber  \\
&& f_8 =(-)^{2j} \frac{ \left[ (2j-1)! \right]}{(4j+1)^2(4j-1)!}
\frac{7\times 5}{16}
\nonumber \\
&& ~~~ \times
\frac{(2j)(2j+1)(2j+2)(2j+3)(2j+4)}{(4j-7)(4j-5)(4j-3)(4j-1)} ~, \nonumber  \\
&& f_{10} = -(-)^{2j} \frac{ \left[ (2j-1)!
\right]}{(4j+1)^2(4j-1)!}
 \frac{9\times 7}{16}
\nonumber \\
&& ~~~ \times
\frac{(2j)(2j+1)\cdots (2j+5)}{(4j-9)(4j-7)\cdots (4j-3)(4j-1)} ~, \nonumber  \\
&& f_{12} =  (-)^{2j} \frac{ \left[ (2j-1)!
\right]}{(4j+1)^2(4j-1)!} \times\frac{11\times 9 \times 7/3}{32}
\nonumber \\
&& ~~~ \times \frac{(2j)(2j+1)\cdots (2j+6)}{(4j-11)(4j-9)\cdots
(4j-3)(4j-1)} ~, \nonumber  \\
&& f_{14} =  -(-)^{2j} \frac{ \left[ (2j-1)!
\right]}{(4j+1)^2(4j-1)!} \frac{13\times 11 \times 9/3}{32}
\nonumber \\
&& ~~~ \times \frac{(2j)(2j+1)\cdots (2j+7)}{(4j-13)(4j-11)\cdots
(4j-3)(4j-1)} ~, \nonumber  \\
&& \nonumber \\&& g_{16} =   (-)^{2j} \frac{ \left[ (2j-1)!
\right]}{(4j+1)^2(4j-1)!} \frac{15\times 13 \times 11 \times
9/3}{256}
\nonumber \\
&& ~~~ \times
\frac{(2j)(2j+1)\cdots (2j+8)}{(4j-15)(4j-13)\cdots (4j-3)(4j-1)} ~, \nonumber  \\
&& f_{18} = -  (-)^{2j} \frac{ \left[ (2j-1)!
\right]}{(4j+1)^2(4j-1)!} \frac{17\times 15 \times 13 \times
11/3}{256}
\nonumber \\
&& ~~~ \times
\frac{(2j)(2j+1)\cdots (2j+9)}{(4j-17)(4j-15)\cdots (4j-3)(4j-1)} ~, \nonumber  \\
&& f_{20} =  (-)^{2j} \frac{ \left[ (2j-1)!
\right]}{(4j+1)^2(4j-1)!} \frac{\frac{19 \times 17\times 15 \times
13 \times 11}{5\times 3}}{512}
\nonumber \\
&& ~~~ \times
\frac{(2j)(2j+1)\cdots (2j+10)}{(4j-19)(4j-17)\cdots (4j-3)(4j-1)} ~, \nonumber  \\
&& f_{22} =  -(-)^{2j} \frac{ \left[ (2j-1)!
\right]}{(4j+1)^2(4j-1)!} \frac{\frac{21 \times 19\times 17 \times
15 \times 13}{5\times 3}}{512}
\nonumber \\
&& ~~~ \times
\frac{(2j)(2j+1)\cdots (2j+11)}{(4j-21)(4j-19)\cdots (4j-3)(4j-1)} ~, \nonumber  \\
&& f_{24} =  (-)^{2j} \frac{ \left[ (2j-1)!
\right]}{(4j+1)^2(4j-1)!} \frac{\frac{23\times 21 \times 19\times
17 \times 15 \times 13}{5\times 3 \times 3}}{512}
\nonumber \\
&& ~~~ \times \frac{(2j)(2j+1)\cdots (2j+12)}{(4j-23)(4j-21)\cdots
(4j-3)(4j-1)} ~. \nonumber
\end{eqnarray}

We define
\begin{eqnarray}
g'_m =  \left\{
     \begin{array}{ccc}
     j    & j  & 2j-1 \\
     j    & j  & 2j-1 \\
     2j -1   & 2j-1  & 4j - m  \end{array}
     \right\}  ~. \nonumber
\end{eqnarray}
$g'_2=g'_4=\frac{1}{2(4j-1)^2}$, see Eq.(5) of Sec. II. For $g'_m$
with larger $m$ we obtain
\begin{eqnarray}
&& g'_6 = -\frac{3(2j-2)(16j-15)}{2(4j-5)(4j-3)^2(4j-1)^2}
 \nonumber \\
&& g'_8 = \frac{15(2j-3)(6j-7)}{2(4j-7)(4j-5)(4j-3)^2(4j-1)^2}
 \nonumber \\
&& g'_{10} =-\frac{7 \times 5 \times 3}{(4j-5)(4j-3)(4j-1)} \nonumber \\
&&  ~~~ \times \frac{(2j-4)(2j-3)(32j-45) }{4(4j-9) (4j-7) \cdots
(4j-1)}
~,  \nonumber \\
&& g'_{12} = \frac{9 \times 7 \times 5 \times 3}{(4j-5)(4j-3)(4j-1)} \nonumber \\
&&  ~~~ \times \frac{(2j+4)(2j-5) (20j-33)}{4(4j-11) (4j-9) \cdots
(4j-1)} ~.  \nonumber
\end{eqnarray}

We define
\begin{eqnarray}
g_I =  \left\{
     \begin{array}{ccc}
     j    & j  & 2j-1 \\
     j    & j  & 2j-1 \\
     2j -1   & 2j-1  &  I  \end{array}
     \right\}  ~. \nonumber
\end{eqnarray}
and obtain
\begin{eqnarray}
&& g_0 = (-)^{2j} \frac{j(4j-3)\left[(2j-1)! \right]^2}{(4j-1)(4j-1)!} ~, \nonumber \\
&& g_2 = -(-)^{2j} \frac{j(8j^2-6j-3) \left[ (2j-1)! \right]^2}{(4j-3)(4j-1)(4j-1)!} ~,  \nonumber \\
&& g_4 = (-)^{2j} \frac{3j(2j+1)(4j^2-3j-5)\left[ (2j-1)! \right]^2}{(4j-5)(4j-3)(4j-1) (4j-1)!} ~,  \nonumber \\
&& g_6 = - (-)^{2j} \frac{j(j+1)(2j+1)\left[ (2j-1)!
\right]^2}{(4j-1)!}  \nonumber \\
&&  ~~~~~  \times  \frac{5 (8j^2-6j-21)}{(4j-7)(4j-5)(4j-3)(4j-1) } ~, \nonumber \\
&& g_8 =  (-)^{2j} \frac{j(j+1)(2j+1)(2j+3)\left[ (2j-1)! \right]^2}{(4j-1)!} \nonumber \\
&&  ~~~~~  \times
\frac{35(4j^2-3j-18)}{2(4j-9)(4j-7)(4j-5)(4j-3)(4j-1) } ~,
\nonumber \\
&& g_{10} = - (-)^{2j} \frac{j(j+1)(j+2)(2j+1)(2j+3)}{(4j-1)!} \nonumber \\
&&  ~~~~~  \times \frac{63(8j^2-6j-55)\left[ (2j-1)!
\right]^2}{2(4j-11)(4j-9) \cdots
 (4j-1) } ~, \nonumber \\
&& g_{12} =   (-)^{2j} \frac{j(j+1)(j+2)(2j+1)(2j+3)(2j+5) }{(4j-1)!} \nonumber \\
&&  ~~~~~  \times \frac{231(4j^2-3j-39) \left[ (2j-1)!
\right]^2}{2(4j-13)(4j-11) \cdots
 (4j-1) } ~   . \nonumber
\end{eqnarray}
Some of above $g_m$ were also obtained for fermions in a
single-$j$ shell in Ref. \cite{Zhao-prc68} where $j$ is a half
integer,  while here $j$ can be either an integer or a half
integer.

\vspace{0.3in}

\begin{center}
{\bf Appendix ~B~ Number of matrices with $K\neq J$ for $I\le 2j$}
\end{center}

Number of matrices  with $J=K$ is always 1 for an even value of
$I$, which contribute $2\times (j+\frac{1}{2})$ on the left hand
side of Eq. (\ref{sum-1}), while that (denoted by $F_J$ here) with
$J\neq K$ is rather complicated.

For $I \le \left[j \right]$ with $J > 2 \left[(I-1) /2\right] $
and $J < 2j - 1  -2\left[\frac{I}{2}\right]$, $F_J = 2
\left[\frac{I}{2}\right]$;

For $I \le \left[j \right]$ with $J < 2 \left[(I-1) /2\right] $,
$F_J = J$;

For $I \le \left[j \right]$ with $J \ge 2j - 1  -2
\left[\frac{I}{2}\right]$, $F_J = \left[\frac{I}{2}\right] +
[\frac{2j-1-J}{2}]$;

For  $\left[j \right] \le I \le 2j$ with $J < 2j-1 -
2\left[\frac{I}{2}\right]$, $F_J = J$;

For  $\left[j \right] \le I \le 2j$ with $J \ge 2j-1 -2\left[I/2
\right]$ and $J <   2 \left[\frac{I}{2}\right] ,  F_J = 2j-1
-2\left[\frac{I}{2}\right] + \left[ \frac{J-(2j-1 -2\left[I/2
\right])}{2} \right]$;

For  $\left[j \right] \le I \le 2j$ with   $J >   2
\left[\frac{I}{2}\right] $, $F_J = (2j-1 - J)/2 +
\left[I/2\right]$.

Because of complexity in the above classification, it is tedious
to show $\sum_{JK} (1+(-)^I)$ by one formula, because one must
simplify many terms such as $\left[ ~ \right]$ which means to take
the largest integer not exceeding the value inside. However, one
can obtain explicit sum rules by writing down their value and
studying their individual modular behavior, as shown in this
paper.

\vspace{0.3in}

\end{document}